\documentclass[10pt,a4]{article}

\usepackage{latexsym,amsmath,amscd,amssymb,graphicx}

\begin{document}
%
\title{The relativistic phase space and Newman-Penrose
basis }

\author{Yaakov Friedman\\
Jerusalem College of
Technology\\ P.O.B. 16031 Jerusalem 91160\\
Israel}
\date{}
\maketitle
\begin{abstract}

We define a complex relativistic phase space which is the space
$\mathbb{C}^4$ equipped with the Minkowski metric and with a
geometric tri-product on it. The geometric tri-product is similar to
the triple product of the bounded symmetric domain of type IV in
Cartan's classification, called the spin domain. We show that there
are to types of tripotents-the basic elements of the tri-product in
the relativistic phase space. We construct a spectral decomposition
for elements of this space. A description of compatibility of
element of the relativistic phase space is given.  We show that the
relativistic phase space has two natural bases consisting of
compatible tripotents. The fist one is the natural basis for
four-vectors and the second one is the Newman-Penrose basis. The
second one determine Dirac bi-spinors on the phase space. Thus, the
relativistic phase space has similar features to the quantum
mechanical state space.
%

\end{abstract}
%
\section{Introduction}

In \cite{F07} we introduced a complex relativistic phase space as
follows. We equip the space $\mathbb{C}^4$  with an  inner product
based on the Lorentz metric and define a new geometric tri-product
on it. This space is used to represent both space-time coordinates
and the four-momentum of an object. The real part of the inner
product extends the notion of an interval, while the imaginary part
extends the symplectic structure of the classical phase space. We
construct both spin 1 and spin 1/2 representations of  the
Poincar\'{e} group by natural operators of the tri-product on the
phase space.

We study now the algebraic properties of the complex relativistic
phase space. We define and describe the tripotents-the basic
elements of the tri-product and show that there are only two types
of such tripotents. The first type will be called maximal tripotents
and the second one minimal ones. We define orthogonality of
tripotents and show that each element can be decomposed as a linear
combination of orthogonal tripotents. This show that the elements of
the relativistic phase space have a decomposition similar to the
decomposition of states in quantum mechanics.

We define the notion of compatibility of tripotents, which is
slightly weaker then the one used in quantum mechanics. We give a
characterization when two tripotents are compatible. We show that
 the relativistic phase space has two natural bases consisting
of compatible tripotents. The fist one is the natural basis for
four-vectors and the second one is the Newman-Penrose basis.

\section{A complex relativistic phase space}

For description of classical motion of a point-like body, the
classical phase space, composed from space position and the
3-momentum, is used. A symplectic structure on the phase space
provides information on the scalar product and the antisymmetric
symplectic form of this space. For the classical phase space this
structure can be expressed efficiently by introducing the complex
structure and a scalar product under which the classical phase space
becomes equal to $\mathbb{C}^3$, see \cite{Kaiser} and
\cite{Arnold}.

 For description of the motion of a \textit{relativistic} point-like body we may use the
relativistic phase space. This is obtained by adding the time and
the energy variables to the classical phase space. So, the
relativistic phase space may be identified with $\mathbb{C}^4,$ in
which the real part expresses the 4-momentum and the imaginary part
represents the space-time position of a point-like body.
 To extend the symplectic structure to this
space, we will use the complex-valued scalar product introduced by
E. Cartan, see \cite{CE66}. Since the scalar product needs to
provide information on the interval of the 4-vectors, we replace the
Euclidian metric, used usually, with the Lorentzian one.

 We define a scalar
product $<\cdot|\cdot>$ on $\mathbb{C}^4$ with natural basic
$\{\mathbf{u}_\mu\}$ as follows:
 For two arbitrary vectors $\mathbf{a},\mathbf{b}\in\mathbb{C}^4$ it is given by
\begin{equation}\label{Minkbilin}
  <\mathbf{a}|\mathbf{b}>=<{a}^\mu \mathbf{u}_\mu|b^\nu
  \mathbf{u}_\nu>=\eta _{\mu \nu} {a}^\mu b^\nu,
\end{equation} where $\mathbf{a}=a^\mu
\mathbf{u}_\mu$ and $\mathbf{b}=b^\mu \mathbf{u}_\mu$. Evidently,
this scalar product is bilinear and symmetric. For an arbitrary
element $\mathbf{a}\in \mathbb{C}^4$, the \textit{scalar square} is
given by
\begin{equation}\label{squaredef}
    \mathbf{a}^2=<\mathbf{a}|\mathbf{a}>=\eta _{\mu \nu} {a}^\mu
    a^\nu ,
\end{equation}
which is a complex number, not necessary positive or even real.

Note that the real subspace $M$ defined by the real vectors of
$\mathbb{C}^4$ with the bilinear form (\ref{squaredef}) may be
identified with the Minkowski space. The same is true for the pure
imaginary subspace $iM$ defined by pure imaginary vectors. We use
the subspace $M$ to represent the four-vector momentum
$\mathbf{p}=p^\mu \mathbf{u}_\mu$ and the subspace $iM$  to
represent the space-time coordinates $i\mathbf{x}=x^\mu i
\mathbf{u}_\mu$ of a point-like body in an inertial system. Thus,
the space $\mathbb{C}^4$ represents both space-time and the
four-momentum as \[\mathbf{a}=a^\mu\mathbf{u}_\mu \;\mbox{with}\;\;
a^\mu=p^\mu+ix^\mu.\]


For any two vectors $\mathbf{a},\mathbf{b}\in \mathbb{C}^4$ we may
consider also a scalar product
$<\overline{\mathbf{a}}|\mathbf{b}>=\eta _{\mu \nu} \overline{a}^\mu
b^\nu.$ The Lorentzian scalar product is the real part
\begin{equation}\label{intervalascomplscalar prod}
    Re<\overline{\mathbf{a}}|\mathbf{b}>=\frac{1}{2}\eta _{\mu \nu}
(\overline{a}^\mu b^\nu+{a}^\mu \overline{b}^\nu),
\end{equation}
 which is symmetric and
 extends the Lorentzian product (and the notion of an interval) from the subspaces $M$ and $iM$ to
$\mathbb{C}^4=M\oplus iM$.

The imaginary part defines a skew scalar product
\begin{equation}\label{liebracL1}
    [\mathbf{a},\mathbf{b}]=Im<\overline{\mathbf{a}}|\mathbf{b}>=
    \frac{1}{2i}\eta _{\mu \nu}
(\overline{a}^\mu b^\nu-{a}^\mu \overline{b}^\nu),
\end{equation}
which extends the symplectic skew scalar product. This bracket can
be used to define the Poisson bracket of two functions and two
vector fields. Thus, the space $\mathbb{C}^4$ with the scalar
product (\ref{Minkbilin}) can be used as a basis for a relativistic
phase space.

 \noindent \textbf{Definition } \textit{Let  $\mathbb{C}^4$ denote a
 4-dimensional complex space with the scalar
product (\ref{Minkbilin}). A }\textbf{geometric tri-product}\textit{
$\{\;,\;,\;\}:\mathbb{C}^4\times\mathbb{C}^4
\times\mathbb{C}^4\rightarrow \mathbb{C}^4$ is defined for any
triple of elements $\mathbf{a}$, $\mathbf{b}$ and $\mathbf{c}$  as
\begin{equation}\label{tripleproddef}
    \mathbf{d}=\{\mathbf{a},{\mathbf{b}},\mathbf{c}\}=<\mathbf{a}
    |{\mathbf{b}}>\mathbf{c}-<{\mathbf{c}}
     |\mathbf{a}> {\mathbf{b}}+<
    {\mathbf{b}}| \mathbf{c}> \mathbf{a}.
\end{equation}
}

In the basis $\{\mathbf{u}_\mu\}$ definition (\ref{tripleproddef})
takes the form
\begin{equation}\label{geomtriprodtensor def}
d^\mu=\eta _{\alpha\beta}a^\alpha {b}^\beta c^\mu-\eta
_{\alpha\beta}c^\alpha a^\beta {b}^\mu+ \eta
_{\alpha\beta}{b}^\alpha c^\beta a^\mu .
\end{equation}

 For any pair of elements
$\mathbf{a},\mathbf{b}\in \mathbb{C}^4$ we define a linear map
$D(\mathbf{a},\mathbf{b}):\mathbb{C}^4\rightarrow\mathbb{C}^4$ as
\begin{equation}\label{Ddef}
   D(\mathbf{a},\mathbf{b})\mathbf{c}=\{\mathbf{a},{\mathbf{b}},\mathbf{c}\}.
\end{equation}
It is easy to verify that the geometric tri-product satisfies the
following properties:

\noindent \textbf{Proposition} The tri-product, defined by
(\ref{tripleproddef}), satisfies:
\begin{enumerate}
 \item $ \{\mathbf{a},{\mathbf{b}},\mathbf{c}\}$ is complex linear in
 all variables
$\mathbf{a},$ $\mathbf{b}$ and $\mathbf{c}.$
\item The triple product is symmetric in the pair of outer variables
\begin{equation}\label{symtriple prod}
   \{\mathbf{a},{\mathbf{b}},\mathbf{c}\}=\{\mathbf{c},{\mathbf{b}},\mathbf{a}\}.
\end{equation}
\item For arbitrary
$\mathbf{x},\mathbf{y},\mathbf{a},\mathbf{b}\in\mathbb{C}^4$,  the
following identity  holds
\begin{equation}\label{dbracket}
 [D(\mathbf{x},\mathbf{y}),D(\mathbf{a},\mathbf{b})]=
D(D(\mathbf{x},\mathbf{y})\mathbf{a},\mathbf{b})
-D(\mathbf{a},D(\mathbf{y},\mathbf{x})\mathbf{b}).
\end{equation}
 \end{enumerate}

Properties of the previous proposition are the defining properties
for the Jordan triple products associated with a homogeneous spaces,
see \cite{L77} and \cite{F04}. If the Euclidean inner product of
$\mathbb{C}^4$ is used  in the definition (\ref{tripleproddef}),
this triple product is the triple product of the bounded symmetric
domain of type IV in Cartan's classification, called the spin
factor. A similar triple product was obtained \cite{F04} for the
ball of relativistically admissible velocities under the action of
the conformal group. As we will see later, the geometric tri-product
(\ref{tripleproddef}) is useful in defining the action od the
Lorentz group on $\mathbb{C}^4$.

   The space $\mathbb{C}^4$
with a form (\ref{Minkbilin}) for the given metric tensor and a
geometric tri-product  (\ref{tripleproddef}) will be  denoted by
$\mathcal{S}^4$. As we have seen, the space $\mathcal{S}^4$ can be
used to represent the space-time coordinates and the relativistic
momentum variables. The form (\ref{Minkbilin}) on it defines both
the interval and the symplectic form and the tri-product may be used
to define the action of the Lorentz group. Thus, we propose to call
$\mathcal{S}^4$  the \textit{complex relativistic phase space}.

\section{Conjugation on $\mathcal{S}^4$}

In order to be able to use effectively the tri-product we will need
to introduce an observer-dependent conjugation on the Relativistic
Phase Space, defined as follows:

 \noindent \textbf{Definition }
\textit{Let $\mathbb{C}^4$ be the complex 4-dimensional space with a
metric tensor $\eta _{\mu\nu}$ on it. For arbitrary element
$\mathbf{b}={b}^\mu \mathbf{u}_\mu$ define a \textbf{conjugate}
$\hat{\mathbf{b}}$ as
\begin{equation}\label{conjugationhat}
    \hat{\mathbf{b}}=\overline{b}^0 \mathbf{u}_0-\overline{{b}}^j\mathbf{u}_j
=\hat{b}^\mu \mathbf{u}_\mu, \; j=1,2,3,
\end{equation}
with $\overline{b}^\mu$ denoting the complex conjugate of $b^\mu$.}

 This conjugation is complex conjugate linear and
combines complex conjugation with space reversal. Since this
conjugation involve space reversal, it is observer dependent and is
not Lorentz invariant. But this is the only deviation from Lorentz
invariance that we will need. This correspond to generally excepted
idea that Quantum Mechanical effects are observer-dependent.

With this conjugation
\begin{equation}\label{Euclidnorm}
   <\hat{\mathbf{b}}|\mathbf{b}>=\sum |b^\mu|^2=|\mathbf{b}|^2
\end{equation}
 coincides with the Euclidian norm in $ \mathbb{C}^4$ and
 \begin{equation}\label{conjugation scalar prod}
   <\hat{\mathbf{a}}|\hat{\mathbf{b}}>=\overline{<\mathbf{a}|\mathbf{b}>}.
 \end{equation}
  It can be
considered as a map to the dual space under the Euclidian norm. Note
that $|\mathbf{b}|^2\geq 0$ and equal to zero only if
$\mathbf{b}=0.$

The conjugation is a triple automorphism, meaning
\begin{equation}\label{conjug triple autom}
    \{\hat{\mathbf{a}},\hat{\mathbf{b}},\hat{\mathbf{c}}\}=\{\mathbf{a},{\mathbf{b}},\mathbf{c}\}\hat{ }.
\end{equation}
By use of this conjugation, for any element $\mathbf{a}\in
\mathcal{S}^4$ we may introduce a linear operator
\begin{equation}\label{Daa}
  D(\mathbf{a})\mathbf{c}= D(\mathbf{a},\hat{\mathbf{a}})\mathbf{c},\;\;\mathbf{c}\in
\mathcal{S}^4.
\end{equation}

\section{Tripotents of the spin space}

The tripotents of the tri-product replace the building blocks for
binary operations - the \textit{idempotents}, defined as non-zero
elements $p$ that satisfy $p^2=p.$
 For a ternary operation,
the building blocks are the \textit{tripotents} defined as:

\noindent \textbf{Definition } A non-zero elements $\mathbf{u}$ of
$\mathcal{S}^4$ is called a \textit{tripotent} if
$\mathbf{u}^{(3)}=\{\mathbf{u},\hat{\mathbf{u}},\mathbf{u}\}=D(\mathbf{u})\mathbf{u}=
\mathbf{u}.$

From the definition of a tripotent  $\mathbf{u}\in {\bf {\cal S}}^4$
and the definition of the triple product (\ref{tripleproddef}) we
get
\begin{equation}\label{tripequation}
   \{\mathbf{u},\hat{\mathbf{u}},\mathbf{u}\}=2 <\hat{\mathbf{u}}|\mathbf{u}>\mathbf{u}
   -\mathbf{u}^2 \hat{\mathbf{u}}=2 |\mathbf{u}|^2\mathbf{u}
   -\mathbf{u}^2 \hat{\mathbf{u}}=
   \mathbf{u},
\end{equation}
or, equivalently,
\begin{equation}\label{tripequation2}
   (2|\mathbf{u}|^2 -1)\mathbf{u}
   - \mathbf{u}^2\hat{\mathbf{u}}=0
   ,\;\Leftrightarrow\; \alpha\mathbf{u}-\beta\hat{\mathbf{u}}=0
\end{equation}
for some constants $\alpha,\beta$.

If  $\alpha\ne 0$, denote $\nu=\beta/\alpha$. Then
$\mathbf{u}=\nu\hat{\mathbf{u}}$. By taking the conjugate of this
equation we get
$\hat{\mathbf{u}}=\overline{\nu}\mathbf{u}=\overline{\nu}\nu\hat{\mathbf{u}}$
 implying $|\nu|=1$. Now
 $|\mathbf{u}|^2=<\hat{\mathbf{u}}|\mathbf{u}>=\overline{\nu}\mathbf{u}^2$ and
 by substituting this in (\ref{tripequation2}) we get
 $\mathbf{u}^2\overline{\nu}\mathbf{u}=\mathbf{u},$ implying
 $\mathbf{u}^2=\nu$ and $|\mathbf{u}|^2=1$.
 Thus, in this
case,
\begin{equation}\label{maxtripotent}
    \mathbf{u}=\nu\hat{\mathbf{u}},\;\;\;
    |\nu|=1, \;\;\;\mathbf{u}^2=\nu,\;\;|\mathbf{u}|^2=1.
\end{equation}
 Tripotents satisfying (\ref{maxtripotent})
will be called \textit{maximal tripotents.} If $\mathbf{u}$  is a
maximal tripotent, then  $D(\mathbf{u})= I$ and
$Q(\mathbf{u})\mathbf{b}=2<\mathbf{u}|\hat{\mathbf{b}}>\mathbf{u}-\nu\hat{\mathbf{b}}$.
Note that our basis vectors $\mathbf{u}_\mu$ are maximal tripotents.

We can obtain an explicit form for maximal tripotents in
$\mathcal{S}^4$. Denoting $\nu=e^{2i\varphi}$ and defining
$\mathbf{a}=e^{-i\varphi}\mathbf{u}=e^{i\varphi}\hat{\mathbf{u}}$ we
obtain $\hat{\mathbf{a}} =e^{i\varphi}\hat{\mathbf{u}}=\mathbf{a}$.
 Decompose $\mathbf{a}$ into the time and spacial components as
$\mathbf{a}=a_0\mathbf{u}_0 +a_j\mathbf{u}_j$ for $j=1,2,3$. Denote
the spacial part $\vec{a}=a_j\mathbf{u}_j$ which can be be written
as $\vec{a}=|\vec{a}|\vec{n}$ with $\vec{n}$ a unit spacial vector
belonging to the sphere
\begin{equation}\label{sphere}
    S_2=\{\hat{n}=n^j\mathbf{u}_j:\;\;n^j\in \mathbb{R},\;
    \sum (n^j)^2=1\}.
\end{equation}
Since $\mathbf{a}=\hat{\mathbf{a}},$ the value $a_0$ is real and
$\vec{a}$ is a purely imaginary vector. Now, $|\mathbf{a}|^2=
|\mathbf{u}|^2=1$ becomes $a_0^2 +|\vec{a}|^2=1.$ This implies that there is a real number $\psi$ and  $\vec{n}\in S_2$ such that $\mathbf{a}=\cos
\psi\mathbf{u}_0 +i\sin \psi\vec{n}.$ Thus, $\mathbf{u}$ is a
maximal tripotent if and only if there are real numbers
$\varphi,\psi$ and $\vec{n}\in S_2$ such that
\begin{equation}\label{maxtripotform}
    \mathbf{u}=e^{i\varphi}\mathbf{a}=e^{i\varphi}(\cos \psi\mathbf{u}_0
+i\sin \psi\vec{n}),\;\;\varphi ,\psi\in \mathbb{R},\;\vec{n}\in
S_2.
\end{equation}
This implies that the collection of all maximal tripotents form a
torus.

If $\alpha=0$, then  from equation (\ref{tripequation2}) we get
$\beta=0$ and thus $\mathbf{u}^2=0$ and $|\mathbf{u}|^2= 1/2.$
Moreover, any element $\mathbf{u}$ satisfying these two properties
will be a tripotent. Such a tripotent will be called a
\textit{minimal tripotent}.
\begin{equation}\label{mintripodef}
    \mathbf{u}-\;\mbox{minimal
    tripotent}\;\Leftrightarrow\;\mathbf{u}^2=0,\;|\mathbf{u}|^2= 1/2.
\end{equation}

Because minimal tripotents $\mathbf{u}$ fulfill $\mathbf{u}^2=0$, they
are called null-vectors in the literature.  Note, though, that not all null-vectors
are minimal tripotents.

 We separate the real and imaginary parts of $\mathbf{u}$ as
$\mathbf{u}=\mathbf{x}+i\mathbf{y}$ and use that
\[\mathbf{u}^2=\mathbf{x}^2-\mathbf{y}^2+2i<{\mathbf{x}}|\mathbf{y}>=0\]
 we get $\mathbf{x}^2=\mathbf{y}^2$ and $<\mathbf{x}|\mathbf{y}>=0$.
  Separating the time and
 space components of the real and imaginary parts of  $\mathbf{u}$
 we get
$<\mathbf{x}|\mathbf{y}>=x_0y_0-<\vec{x}|\vec{y}>=0.$ If we denote
$\lambda=\mathbf{x}^2=\mathbf{y}^2$ then $x_0^2=|\vec{x}|^2+\lambda$
and similarly
 $y_0^2=|\vec{y}|^2+\lambda.$ But if $\lambda\geq 0$
 then
 \[(x_0y_0)^2=|<\vec{x}|\vec{y}>|^2=(|\vec{x}|^2+\lambda)
 (|\vec{y}|^2+\lambda)\geq|\vec{x}|^2|\vec{y}|^2.\]

  The Cauchy-Schwartz inequality for vectors in $\mathbb{R}^3$ excludes $\lambda >0$
  implying that both $\mathbf{x}$ and $\mathbf{y}$ are
space-like or light-like elements of $M$. If $\mathbf{x}$ and
$\mathbf{y}$ are light-like, then from the equality in the
 Cauchy-Schwartz inequality follow that the vectors $\vec{x}$ and $\vec{y}$ are collinear.
From $<\mathbf{x}|\mathbf{y}>=0$ and $\mathbf{x}^2=\mathbf{y}^2 = 0$
follow that $\mathbf{y}=\alpha\mathbf{x}$ for some constant
$\alpha$. Since a minimal tripotent $\mathbf{u}$ satisfy
$|\mathbf{u}|^2=1/2$ there are a real constant $\psi$ and
$\vec{n}\in S_2$ such that
\begin{equation}\label{nulltripotform}
    \mathbf{u}=\frac{e^{i\psi}}{2}(\mathbf{u}_0
+\vec{n}),\;\;\psi \in \mathbb{R},\;\vec{n}\in S_2.
\end{equation}
An arbitrary minimal tripotent satisfies
\begin{equation}\label{mintripot}
   \mathbf{u}=\mathbf{x}+i\mathbf{y},\:\;\mathbf{x},\mathbf{y}\in
   M,\;\;
   \mathbf{x}^2=\mathbf{y}^2 \leq 0,\;\;<{\mathbf{x}}|\mathbf{y}>=0.
 \end{equation}

If $\mathbf{u}$ is a minimal tripotent, the operator
$D(\mathbf{u})$, defined by (\ref{Daa}) acts as follows:
\begin{equation}\label{Dumintripo}
   D(\mathbf{u})\mathbf{a}=\frac{1}{2}\mathbf{a}-<\mathbf{u}|\mathbf{a}>\hat{\mathbf{u}}+
<\hat{\mathbf{u}}|\mathbf{a}>\mathbf{u}.
\end{equation}
It follows from this that if $\mathbf{a}$ is an eigenvector of
$D(\mathbf{u})$ corresponding to eigenvalue $\alpha$,
 then:
 \begin{equation}\label{Dmintripo}
   D(\mathbf{u})\mathbf{a}=\alpha\mathbf{a},\; \alpha=\left\{
             \begin{array}{ll}
               1, & \hbox{ $\mathbf{a}=\lambda\mathbf{u},\;\lambda\in\mathbb{C}$} \\
               1/2, & \hbox{ $<\mathbf{u}|\mathbf{a}>=0$ and $<\hat{\mathbf{u}}|\mathbf{a}>=0$} \\
               0, & \hbox{ $\mathbf{a}=\lambda\hat{\mathbf{u}},\;\lambda\in\mathbb{C}$}
             \end{array}
           \right .
\end{equation}
 This implies that if $\mathbf{u}$ is a minimal tripotent then the spectrum of the operator $D(\mathbf{u})$ is the set  $\{1,1/2,0\}$.

 The following table summarizes the properties of the two types of
tripotents in $\mathcal{S}^4$.

\medskip

\noindent \begin{tabular}{|c|c|c|c|c|c|}
                                \hline
                                                                Type & $|\mathbf{u}^2|$ & $|\mathbf{u}|^2$ & sp$D(\mathbf{u})$ &
                                $Q(\mathbf{u})\mathbf{a}$ & Decomposition \\
                                \hline
                                Maximal & 1 & 1 & 1 & $(2P_ \mathbf{u}-\nu){\mathbf{a}}$&
                                 (\ref{maxtripotform}) \\
                                Minimal & 0 & $\frac{1}{2}$ &$\{1,\frac{1}{2},0\}$ & $2P_ \mathbf{u}{\mathbf{a}}$ &
                                (\ref{mintripot})\\
                                \hline
                              \end{tabular}

 {\centerline{ Table 1. Types of tripotents in $\mathcal{S}^4$.}}

\medskip
The operator $P_ \mathbf{u}$ in the table is defined by $P_
\mathbf{u}\mathbf{b}=<\mathbf{u}|{\mathbf{b}}>\mathbf{u}$.

\section{Orthogonality of tripotents and Spectral decomposition}

 \noindent \textbf{Definition } Let  $\mathbf{u}$ be a tripotent in
 $\mathcal{S}^4$. We will say that a tripotent $\mathbf{w}$ is
 \textit{orthogonal (or algebraically orthogonal)} to $\mathbf{u}$
 if
  \begin{equation}\label{orthogtripot}
  D(\mathbf{u})\mathbf{w}=0.\end{equation}

 If $\mathbf{u}$ is a maximal tripotent, then
 $D(\mathbf{u})=I$ and there are no tripotents orthogonal to $\mathbf{u}.$
 From the definition of a minimal tripotent it follows that if $\mathbf{u}$
 is a minimal tripotent,  $\hat{\mathbf{u}}$ is also a minimal
 tripotent and from (\ref{Dmintripo}) it follows that  $D(\mathbf{u})\hat{\mathbf{u}}=0.$
  So, if $\mathbf{u}$ is a
 minimal tripotent, the tripotent $\hat{\mathbf{u}}$ is orthogonal
to $\mathbf{u}$.

 Let  $\mathbf{w}$ be an arbitrary tripotent orthogonal to a minimal tripotent
$\mathbf{u}.$ From the definition of orthogonality and
(\ref{Dmintripo}) we get $\mathbf{w}=\lambda\hat{\mathbf{u}}.$
 From (\ref{mintripodef}) it follows that $|\lambda|=1$.
 Conversely, if
$\mathbf{w}=\lambda \hat{\mathbf{u}}$ then from (\ref{Dmintripo}) $
D(\mathbf{u})\mathbf{w}=0$ implying that $\mathbf{w}$ is orthogonal
to $\mathbf{u}.$ Thus
\begin{equation}\label{orthogonalform}
 \mathbf{w}\mbox{  is orthogonal to  } \mathbf{u}\;\Leftrightarrow\;
\mathbf{w}=\lambda
\hat{\mathbf{u}},\;\lambda\in\mathbb{C},\;|\lambda|=1
\end{equation}

Since $\mathbf{w}=\lambda \hat{\mathbf{u}}$ imply that
$\mathbf{u}=\lambda \hat{\mathbf{w}},$ it follows that $\mathbf{w}$
is orthogonal to $\mathbf{u}$ if and only if $\mathbf{u}$ is
orthogonal to $\mathbf{w}.$ We will call such pair of tripotents an
\textit{orthogonal pair} and will denote this
$\mathbf{u}\perp\mathbf{w}.$ If $\mathbf{u}\perp\mathbf{w},$  then
from (\ref{Ddef}) and (\ref{mintripodef}) follow that
 \begin{equation}\label{orthogtripot2}
  \mathbf{u}\perp\mathbf{w}\;\Leftrightarrow\;D(\mathbf{u},\hat{\mathbf{w}})
  =D(\mathbf{w},\hat{\mathbf{u}})=0.\end{equation}
Moreover if
\begin{equation}\label{decomposorthog}
   \mathbf{a}=\alpha\mathbf{u}+\beta\mathbf{w}
\end{equation}
for some constants $\alpha ,\beta$ then
$D(\mathbf{a})=|\alpha|^2D(\mathbf{u})+|\beta |^2D(\mathbf{w})$ and
thus
\begin{equation}\label{qubeinspectral}
   \mathbf{a}^{(3)}=D(\mathbf{a})\mathbf{a}=|\alpha|^2\alpha\mathbf{u}
+|\beta|^2\beta\mathbf{w}.
\end{equation}

This shows that orthogonal tripotents behave as two orthogonal
projections and hence that $\mathbf{u}+\mathbf{w}$ and
$\mathbf{u}-\mathbf{w}$ are maximal tripotents.
We may now introduce a partial order on the set of tripotents. We say
that $\mathbf{u}<\mathbf{v}$ for a pair of tripotents
$\mathbf{u},\mathbf{v}$ if there is a tripotent $\mathbf{w}$
orthogonal to $\mathbf{u}$ such that
$\mathbf{v}=\mathbf{u}+\mathbf{w}.$ Under such an ordering the
minimal tripotents are minimal and the maximal ones are maximal.

 If $\mathbf{a}$ has decomposition
 (\ref{decomposorthog}) and $\mathbf{w}=\lambda
\hat{\mathbf{u}}$, then $\mathbf{a}^2= \lambda\alpha\beta$.  This shows
that $\mathbf{a}^2$ has the meaning of a determinant. If $\mathbf{a}^2=0$ then  it follows from
(\ref{mintripodef}) that
$\mathbf{u}=\frac{\mathbf{a}}{2|\mathbf{a}|}$ is a minimal tripotent.  This implies, in turn, that $\mathbf{a}=2|\mathbf{a}|\mathbf{u}$ has a
decomposition (\ref{decomposorthog}) with a positive constant
$\alpha =2|\mathbf{a}|$. The following Proposition shows that for any
element in $\mathcal{S}^4$ there is such a decomposition with positive
constants.

\noindent \textbf{Proposition} Let  $\mathbf{a}$ be a non-zero
element of $\mathcal{S}^4$ with $\mathbf{a}^2\ne 0$. Then there is
an orthogonal pair of minimal tripotents $\mathbf{u}$ and
$\mathbf{w}$ such that
\begin{equation}\label{spectraldecomp}
    \mathbf{a}=s_1\mathbf{u}+s_2\mathbf{w},
\end{equation}
where the pair of non-negative numbers  $s_1,s_2$ (called singular
numbers of $\mathbf{a}$) can be defined from the equations
\begin{equation}\label{singularnumdef}
   s_1\pm s_2=\sqrt{2|\mathbf{a}|^2\pm
   2|\mathbf{a}^2|}.
\end{equation}
  Moreover, if $s_1> s_2$ this
decomposition is unique.

Proof.

Use the polar decomposition of complex numbers to decompose
$\mathbf{a}^2=\lambda|\mathbf{a}^2|$ with $|\lambda |=1.$ Since
$\hat{\mathbf{a}}^2=\overline{\mathbf{a}^2}$ we have
$\hat{\mathbf{a}}^2=\bar{\lambda}|\mathbf{a}^2|.$
 Define $\mathbf{b}=\mathbf{a}+\lambda\hat{\mathbf{a}}$. Then
 $\hat{\mathbf{b}}=\bar{\lambda}\mathbf{b}$ and
 $|\mathbf{b}|^2=\bar{\lambda}\mathbf{b}^2=2|\mathbf{a}|^2+
   2|\mathbf{a}^2|.$
Similarly define $\mathbf{c}=\mathbf{a}-\lambda\hat{\mathbf{a}}$.
Then
 $\hat{\mathbf{c}}=-\bar{\lambda}\mathbf{c}$ and
 $|\mathbf{c}|^2=-\bar{\lambda}\mathbf{c}^2=2|\mathbf{a}|^2-
   2|\mathbf{a}^2|.$ Direct calculation shows that
   $<\mathbf{b}|\mathbf{c}>=<\hat{\mathbf{b}}|\mathbf{c}>
   =<\mathbf{b}|\hat{\mathbf{c}}>=0.$ Obviously,
   $\mathbf{a}=\frac{1}{2}(\mathbf{b}+\mathbf{c})$.

   Consider first the case $\mathbf{c}=0$ or $\mathbf{a}=\lambda\hat{\mathbf{a}}$.
For such $\mathbf{a}$ the element
$\mathbf{v}={\mathbf{a}}/{|\mathbf{a}|}$  is
a maximal tripotent as per (\ref{maxtripotent}) and thus, according to (\ref{maxtripotform}), has the form
$\mathbf{v}=e^{i\varphi}(\cos \psi\mathbf{u}_0 +i\sin \psi\vec{n})$
with $\varphi ,\psi\in \mathbb{R}$ and a unit vector
$\vec{n}\in\mathbb{R}^3.$ If we define
\[\mathbf{u}=\frac{e^{i(\varphi+\psi)}}{2}( \mathbf{u}_0 +\vec{n}),\;
\mathbf{w}=\frac{e^{i(\varphi-\psi )}}{2}( \mathbf{u}_0 -\vec{n}),\]
then from (\ref{nulltripotform}) and (\ref{orthogonalform}) it follows
that both $\mathbf{u},\mathbf{w}$ are minimal orthogonal tripotents
and
$\mathbf{a}=|\mathbf{a}|\mathbf{v}=|\mathbf{a}|(\mathbf{u}+\mathbf{w}).$
So, in this case $\mathbf{a}$ has decomposition
(\ref{spectraldecomp}).

Now consider the case $\mathbf{c}\ne 0$. Note that $\mathbf{b}\ne 0$
always. Define
\[\mathbf{u}=\frac{1}{2}\left(\frac{\mathbf{b}}{|\mathbf{b}|}
+\frac{\mathbf{c}}{|\mathbf{c}|}\right),\;\;\mbox{and}\;\;
\mathbf{w}=\frac{1}{2}\left(\frac{\mathbf{b}}{|\mathbf{b}|}
-\frac{\mathbf{c}}{|\mathbf{c}|}\right).\] From the properties of
$\mathbf{b}$ and $\mathbf{c}$ follows that
$|\mathbf{u}|^2=|\mathbf{w}|^2=1/2$ and
$\mathbf{u}^2=\mathbf{w}^2=0$. Thus from (\ref{mintripodef}) it follows
that both $\mathbf{u}$ and $\mathbf{w}$ are minimal tripotents and
since $\hat{\mathbf{u}}=\bar{\lambda}\mathbf{w}$ they are also
orthogonal. The expression
\[\mathbf{a}=\frac{1}{2}(\mathbf{b}+\mathbf{c})=\frac{|\mathbf{b}|+|\mathbf{c}|}{2}
\mathbf{u}+\frac{|\mathbf{b}|-|\mathbf{c}|}{2} \mathbf{w}\] also defines a
decomposition (\ref{spectraldecomp}) for $\mathbf{a}$ in this case
with
\[s_1=\frac{|\mathbf{b}|+|\mathbf{c}|}{2},\;\;s_2=\frac{|\mathbf{b}|-|\mathbf{c}|}{2}\]
being non-negative numbers satisfying (\ref{singularnumdef}) and
$s_1> s_2$.

 To show the uniqueness of the decomposition in the case $s_1\neq s_2$, assume
 that a given $\mathbf{a}$ has a decomposition (\ref{spectraldecomp}) with an
orthogonal pair of minimal tripotents $\mathbf{u}$ and $\mathbf{w}$.
From (\ref{orthogonalform}) we may assume that $\mathbf{w}=
\lambda\hat{\mathbf{u}}$ with $|\lambda|=1$ and by use of
(\ref{mintripodef}) we get
\[|\mathbf{a}|^2=<s_1\hat{\mathbf{u}}+
s_2\bar{\lambda}\mathbf{u}|s_1\mathbf{u}+s_2\lambda\hat{\mathbf{u}}>
=\frac{1}{2}(s_1^2+s_2^2)\] and
\[\mathbf{a}^2=<s_1\mathbf{u}+s_2\lambda\hat{\mathbf{u}}
|s_1\mathbf{u}+s_2\lambda\hat{\mathbf{u}}> =\lambda s_1s_2.\] This
implies equation (\ref{singularnumdef}) for $s_1,s_2$ and that these
numbers are defined uniquely. Note that $\lambda $ is the argument
of the complex number $\mathbf{a}^2$ which is also defined uniquely.
Denote $\mathbf{d}=\mathbf{a}/s_1=\mathbf{u}+\alpha\mathbf{w}$ with
$1>\alpha$. Then from (\ref{qubeinspectral}) we have
\[\lim_{n\mapsto\infty}D(\mathbf{d})^n\mathbf{d}=\lim_{n\mapsto\infty}\mathbf{u}
+\alpha^{2n}\mathbf{w}=\mathbf{u}\] implying that $\mathbf{u}$ is
defined uniquely and therefore the decomposition (\ref{spectraldecomp})
is unique.

\section{Compatible tripotents of the spin space}

From Quantum Mechanics we know that it is preferable to work with a
basis consisting of compatible observables. Two observables are said
to be compatible if the result of measurement of one of them is not
affected by the measurement of the second one. Compatibility mean
commutativity of spectral projections. For self-adjoint operators
this is equivalent to commutativity of the operators representing
the observables. To define compatibility for a pair of elements
$\mathbf{a},\mathbf{b}$ in the spin space $\mathcal{S}^4,$ we use
the fact that there are two operators
$D(\mathbf{a},\hat{\mathbf{a}}),{Q}(\mathbf{a})$ on $\mathcal{S}^4 $
associated with any element $\mathbf{a}\in \mathcal{S}^4.$
Commutativity of spectral projections of two elements in
$\mathcal{S}^4 $ can be replaced with commutativity of the operators
associated with them.

\noindent \textbf{Definition } For any element $\mathbf{a}\in
\mathcal{S}^4 $ we denote two operators
$G_1(\mathbf{a})=D(\mathbf{a},\hat{\mathbf{a}})$ and
$G_2(\mathbf{a})=Q(\mathbf{a})$. A pair of non-zero elements
$\mathbf{a},\mathbf{b}\in \mathcal{S}^4 $ is said to be
\textit{compatible} if
\[[G_j(\mathbf{a}),G_k(\mathbf{b})]=0\;\hbox{for any}\;j,k\in\{1,2\}.\]

For any linearly independent pair of tripotents $
\mathbf{u},\mathbf{v}$ from Table 1 follow that the following are
equivalent:
\[[{Q}(\mathbf{u}),{Q}(\mathbf{v})]=0\;\Leftrightarrow\;
[P_\mathbf{u},P_\mathbf{v}]=0\;\Leftrightarrow\;<\hat{\mathbf{u}}|\mathbf{v}>=0,\]
with $P_ \mathbf{u}\mathbf{b}=<\mathbf{u}|{\mathbf{b}}>\mathbf{u}$.
Thus any pair of linearly independent compatible tripotents $
\mathbf{u},\mathbf{v}$ satisfy $<\hat{\mathbf{u}}|\mathbf{v}>=0.$
The following Proposition describes when a pair of tripotents are
compatible.

\noindent \textbf{Proposition} Let $ \mathbf{u},\mathbf{v}$ be a
pair of linearly independent tripotents in $\mathcal{S}^4.$  The
pair is compatible if and only if: $<\hat{\mathbf{u}}|\mathbf{v}>=0$
and, in addition, if one of the tripotents, say $\mathbf{u},$ is a
minimal one then
 the other one $\mathbf{v}$  must be an eigenvector
of $D(\mathbf{u})$ corresponding to eigenvalue 0  or  $ 1/2$.

 Proof.

Consider first the case in which neither of the tripotents $ \mathbf{u}$ and
$\mathbf{v}$ are minimal. Thus, each of the tripotents $ \mathbf{u}$
and $\mathbf{v}$ is  a maximal  tripotent. Since $D(\mathbf{u})=I$
for a maximal tripotent,  any commutator with $D(\mathbf{u})$ or
with $D(\mathbf{v})$ will vanish. Thus compatibility in such case is
equivalent to $[{Q}(\mathbf{u}),{Q}(\mathbf{v})]=0$ which is
equivalent to the condition $<\hat{\mathbf{u}}|\mathbf{v}>=0$.

We may assume that  $\mathbf{u}$ is a minimal tripotent and
$\mathbf{v}$ is a tripotent compatible with $\mathbf{u}$ satisfying
$<\hat{\mathbf{u}}|\mathbf{v}>=0$. In this case
$D(\mathbf{u})\mathbf{a}$ was defined by (\ref{Dumintripo}) and
$Q(\mathbf{v})\mathbf{a}=(2P_\mathbf{u}+\lambda
I)\mathbf{a}=<\mathbf{u}|\hat{\mathbf{a}}>{\mathbf{u}}.$ Thus,
$[{Q}(\mathbf{v}),D(\mathbf{u})]=0$ if and only if
$[P_\mathbf{v},D(\mathbf{u})]=0$. But
\[P_\mathbf{v}D(\mathbf{u})\mathbf{a}=\frac{1}{2}
<\mathbf{v}|\hat{\mathbf{a}}>{\mathbf{v}}+
<\mathbf{u}|\hat{\mathbf{a}}>
<\hat{\mathbf{u}}|\mathbf{v}>\hat{\mathbf{v}}
-<\hat{\mathbf{u}}|\hat{\mathbf{a}}><\mathbf{v}|\mathbf{u}>
{\mathbf{v}}\]
\[=
\frac{1}{2} <\mathbf{v}|\hat{\mathbf{a}}>{\mathbf{v}}-
<\hat{\mathbf{u}}|\hat{\mathbf{a}}><\mathbf{v}|\mathbf{u}>
{\mathbf{v}}.\]
 A similar calculation will give
\[D(\mathbf{u})P_\mathbf{v}\mathbf{a}=\frac{1}{2}
<\mathbf{v}|\hat{\mathbf{a}}>{\mathbf{v}}
-<\mathbf{v}|\hat{\mathbf{a}}>
<\mathbf{u}|\mathbf{v}>\hat{\mathbf{u}}.\]
 Thus,
$[Q(\mathbf{v}),D(\mathbf{u})]=0$ in the following 2 cases 1)
$\mathbf{v}=\hat{\mathbf{u}}$ or 2) $<\mathbf{u}|{\mathbf{v}}>=0$.

In case 1), from (\ref{Dmintripo}) it follows that $\mathbf{v}$ is
an eigenvector of $D(\mathbf{u})$ corresponding to eigenvalue 0 and
is a minimal tripotent orthogonal to $\mathbf{u}$. In case 2)
$\mathbf{v}$ is an eigenvector of $D(\mathbf{u})$ corresponding to
eigenvalue $1/2$.

Conversely, if $\mathbf{u}$ is a minimal  tripotent and  tripotent
$\mathbf{v}$  is  an eigenvector of $D(\mathbf{u})$ corresponding to
eigenvalue 0, then from (\ref{Dmintripo}) it follows that
$\mathbf{v}=\lambda \hat{\mathbf{u}}$ for some constant $\lambda$.
From the above observations
$[Q(\mathbf{v}),D(\mathbf{u})]=[Q(\mathbf{u}),D(\mathbf{v})]=
[Q(\mathbf{v}),Q(\mathbf{u})]=0$.  Also in this case we get
\begin{equation}\label{DDorthogtrip}
   D(\mathbf{v})D(\mathbf{u})\mathbf{a}=\frac{1}{2}(\frac{1}{2}\mathbf{a}
-<\mathbf{u}|\mathbf{a}>\mathbf{v}-<\mathbf{v}|\mathbf{a}>\mathbf{u})=
D(\mathbf{u})D(\mathbf{v})\mathbf{a},
\end{equation}
based on (\ref{Dumintripo}) and implying that both $[D(\mathbf{v})D(\mathbf{u})]=0$ and that $\mathbf{v}$
and $\mathbf{u}$ are compatible.

If $\mathbf{u}$ is a minimal  tripotent  and $\mathbf{v}$ satisfying
$<\hat{\mathbf{u}}|\mathbf{v}>=0$ is an eigenvector of
$D(\mathbf{u})$ corresponding to eigenvalue $1/2$, then from
(\ref{Dmintripo}) it follows that $<\mathbf{u}|{\mathbf{v}}>=0.$
From the above observations $[Q(\mathbf{v}),D(\mathbf{u})]=0$. Since
for a maximal tripotent $D(\mathbf{v})=I$, in case the tripotent
$\mathbf{v}$ is maximal, the tripotents $\mathbf{v}$ and
$\mathbf{u}$ are compatible.

It remains to consider the case in which $\mathbf{v}$ is  a minimal tripotent.
Such a pair of tripotents $(\mathbf{u},\mathbf{v})$ is called a
\textit{co-orthogonal} pair of tripotents, which is denoted by
$\mathbf{u}\perp\mathbf{v}$. From (\ref{Dumintripo}) we get
\[
D(\mathbf{v})D(\mathbf{u})\mathbf{a}=\]\[\frac{1}{4}\mathbf{a}
+\frac{1}{2}<\hat{\mathbf{u}}|\mathbf{a}>\mathbf{u}+
\frac{1}{2}<\hat{\mathbf{v}}|\mathbf{a}>\mathbf{v}
-\frac{1}{2}<\mathbf{u}|\mathbf{a}>\hat{\mathbf{u}}-
\frac{1}{2}<\mathbf{v}|\mathbf{a}>\hat{\mathbf{v}}\]\[=
D(\mathbf{u})D(\mathbf{v})\mathbf{a},\]implying that
$[D(\mathbf{v})D(\mathbf{u})]=0$ and that $\mathbf{v}$ and
$\mathbf{u}$ are compatible in this case, as well.

\section{Bases of the relativistic phase space}

 The natural basis
$\mathbf{u}_\mu$ of the relativistic phase space $\mathcal{S}^4$
consists of a family of maximal tripotents satisfying
$<\hat{\mathbf{u}}_\mu |\mathbf{u}_\nu>=0$ for any indices $\mu \ne
\nu$. Thus, from Proposition 6.1 it follows that the natural basis
consists a family of compatible maximal tripotents.

 With the spin triple product based on the Euclidean
metric  we had a set of bases, called the \textit{spin grid,} constructed from a family of
compatible minimal tripotents, in addition to the bases from maximal tripotents- see \cite{F04} p.121-123 and
\cite{DF87}. If the spin space is of dimension 4, the spin grid is
also called  an \textit{odd quadrangle}.

\noindent \textbf{Definition } A family of 4 minimal tripotents
$(\mathbf{v},\mathbf{w},\tilde{\mathbf{v}},\tilde{\mathbf{w}})$ in
$\mathcal{S}^4$ is called a \textit{ spin grid} if
\begin{enumerate}
          \item the pairs $(\mathbf{v},\tilde{\mathbf{v}})$ and
$(\mathbf{w},\tilde{\mathbf{w}})$ are orthogonal
\item the pairs $(\mathbf{v},\mathbf{w}),(\mathbf{w},\tilde{\mathbf{v}}),
(\tilde{\mathbf{v}},\tilde{\mathbf{w}}),(\tilde{\mathbf{w}},{\mathbf{v}})$
are co-orthogonal
 \item
$2\{\mathbf{v},\mathbf{w},\tilde{\mathbf{v}}\}=-\tilde{\mathbf{w}}$
and
$2\{\mathbf{w},\tilde{\mathbf{v}},\tilde{\mathbf{w}}\}=-{\mathbf{v}}.$
                         \end{enumerate}

 The spin grid of the relativistic phase
 space $\mathcal{S}^4$ can be constructed as follows.
Start with a minimal light-like tripotent defined by
(\ref{nulltripotform}) as
$\mathbf{v}=\frac{1}{2}(\mathbf{u}_0+\mathbf{u}_3)$. The tripotent
$\hat{\mathbf{v}}=\frac{1}{2}(\mathbf{u}_0-\mathbf{u}_3)$ is
orthogonal to it. From Proposition 6.1 it follows that in order for
the remaining two basic tripotents to be compatible with
 $\mathbf{v}$ and $\hat{\mathbf{v}}$ they must be $-1/2$ eigenvectors of the
 operators $D(\mathbf{v})$ and $D(\hat{\mathbf{v}})$.
 From (\ref{Dmintripo}) it follows that such a tripotent, say $\mathbf{z}$, must
satisfy $<\mathbf{z}|\mathbf{v}>=0$ and
$<\mathbf{z}|\hat{\mathbf{v}}>=0.$ Thus, such
  $\mathbf{z}$ must belong to the subspace $X= span
\{\mathbf{u}_1,\mathbf{u}_2\}$. In this subspace we do not have
light-like minimal tripotents. But, by (\ref{mintripot})
$\mathbf{w}=\frac{1}{2}(\mathbf{u}_1+i\mathbf{u}_2)\in X$  is a
minimal tripotent and
 its orthogonal complement $\hat{\mathbf{w}}=
 \frac{1}{2}(-\mathbf{u}_1+i\mathbf{u}_2)$
 is also a minimal tripotent in $X$. Direct calculations show that
the family of 4 minimal tripotents
$(\mathbf{v},\mathbf{w},\hat{\mathbf{v}},\hat{\mathbf{w}})$ form a
spin grid.

 Such
a basis is the known Newman-Penrose basis (see
\cite{PenroseRindler})
$(\mathbf{l},\mathbf{m},\overline{\mathbf{m}},\mathbf{n})$ with
\[\mathbf{l}=\frac{1}{2}(\mathbf{u}_0+\mathbf{u}_3),\;\;\mathbf{m}=\frac{1}{2}(\mathbf{u}_1+i\mathbf{u}_2)\]
\begin{equation}\label{neumanPenroseBasis}
  \overline{\mathbf{m}}=\frac{1}{2}(\mathbf{u}_1-i\mathbf{u}_2),\;\;\mathbf{n}=\frac{1}{2}(\mathbf{u}_0-\mathbf{u}_3).
\end{equation}

The spin grid differ from the NP basis in 1) the normalization
constant, 2) the order of the basis elements, which was not critical
for the NP basis, and 3) the minus in front of
$\overline{\mathbf{m}},$ which is needed to be able to extend such a
basis to higher dimensions (see \cite{F04}, p.245). The minus in
front of $\overline{\mathbf{m}}$ has been added in consideration of
the symmetries of spinors (see \cite{ODonnell}). Thus, we propose to
call spin grid a modified NP basis.

\noindent{\textbf{Definition }}A family of 4 minimal tripotents
$(\mathbf{v},\mathbf{w},\hat{\mathbf{v}},\hat{\mathbf{w}})$ in
$\mathcal{S}^4$ is called a \textit{ Modified Newman-Penrose} basis
if
\[\mathbf{v}=\frac{1}{2}(\mathbf{u}_0+\mathbf{u}_3),\;\;\mathbf{w}=
\frac{1}{{2}}(\mathbf{u}_1+i\mathbf{u}_2)\]

\begin{equation}\label{ModifiedneumanPenroseBasis}
  \hat{\mathbf{v}}=\frac{1}{{2}}(\mathbf{u}_0-\mathbf{u}_3),\;\;\hat{\mathbf{w}}
  =\frac{1}{{2}}(-\mathbf{u}_1+i\mathbf{u}_2).
\end{equation}
 Such a basis form a spin grid.

\medskip

I want to thank  Yakov Itin for helpful remarks and suggestions.


\begin{thebibliography}{95}

\bibitem{Arnold} V. I. Arnold, \textit{Mathematical Methods of Classical
Mechanics}, Springer-Verlag, New York, 1978.


\bibitem{CE35} E. Cartan, Sur les domains born\'{e}s homog\`{e}nes de
l'espace de $n$ variables complexes, \textit{ Abh. Math. Sem.
Univ. Hamburg }\textbf{11} (1935), 116--162.

\bibitem{CE66} E. Cartan, \textit{The Theory of Spinors,} Dover
Publ. New York, 1966.

\bibitem{DF87} T. Dang and Y. Friedman,  Classification of $JBW^*$-triples and applications,
\textit{Math. Scand.} \textbf{61} (1987), 292--330.

\bibitem{F04}Y. Friedman, \textit{Physical Applications of Homogeneous Balls},
Progress in Mathematical Physics \textbf{40} (Birkh\"{a}user,
Boston, (2004).

\bibitem{F07}Y. Friedman, \textit{Representations of the Poincar\'{e} group on relativistic phase space},
 arXiv:0802.0070v1, 2008.

%
%

\bibitem{FR01} Y. Friedman, B. Russo, A new approach to spinors
and some representation of the Lorentz group on them,
\textit{Foundations of Physics}, \textbf{31}(12), (2001),
1733--1766.

%

 \bibitem{Kaiser} G. Kaiser, \textit{Quantum Physics, Relativity, and
 Complex Spacetime}, North-Holland, Amsterdam, 1990.

 \bibitem{L77}  O. Loos, \textit{Bounded
symmetric domains and Jordan pairs}, University of
 California, Irvine, 1977.

  \bibitem{ODonnell} P. O'Donnell, \textit{Introduction to 2-Spinors in
  General Relativity}, World Scientific Pub., 2003.

 \bibitem{PenroseRindler} R. Penrose \& W. Rindler, \textit{Spinors and
 space-time}, Cabridge University Press, 1984.

%

\end{thebibliography}
\end{document}